\documentclass{webofc}

\usepackage[varg]{txfonts}   
\usepackage{hyperref}
\usepackage{url}
\hypersetup{colorlinks=true,citecolor=blue,urlcolor=blue,linkcolor=blue}
\begin{document}

\title{Tomography of light hadron structures: From LCDAs to iTMDs}

\author{
\firstname{Shan} \lastname{Cheng}\inst{1,2}\fnsep
\thanks{\email{scheng@hnu.edu.cn}, speaker} \and
\firstname{Jian} \lastname{Chai}\inst{3}\fnsep
\and
\firstname{Ji-Xin} \lastname{Yu}\inst{4}\fnsep
}

\institute{School of Physics and Electronics, Hunan University, 410082 Changsha, China, 
\and
School for Theoretical Physics and Hunan Provincial Key Laboratory of High-Energy Scale Physics and Applications, Hunan University, Changsha 410082, China
\and
School of Physics and Advanced Energy, Henan University of Technology, Zhengzhou, Henan 450001, China
\and 
Frontiers Science Center for Rare Isotopes and School of Nuclear Science and Technology, Lanzhou University, Lanzhou 730000, China
          }

\abstract{To account for the soft transverse dynamics overlooked by the $k_T$ factorization formalism of exclusive QCD processes, we propose introducing the intrinsic transversal momentum distribution functions (iTMDs), in conjunction with the light-cone distribution amplitudes (LCDAs), to define the probability amplitude for finding a hadron state where partons move fast along the longitudinal axis while undergoing gentle oscillations in the transverse plane. 
We implement this framework in the calculation of the electromagnetic form factors of the pion and the proton. Our iTMDs-improved perturbative QCD predictions show good agreement with available experimental data and lattice QCD results. More importantly, we find that higher-twist, especially subleading-twist contributions, are indispensable for explaining the scaling behavior of both pion and proton form factors at intermediate momentum transfers. Furthermore, the inclusion of iTMD effects significantly improves the perturbative QCD predictions for the pion form factor in the few GeV$^2$.}

\maketitle

\section{Introduction} \label{intro}  

In particle physics, the hadron form factor is a fundamental observable that encapsulates the properties of an interaction without resolving the hadron's internal breakup \cite{Mueller:1981sg,PQCD}. As a dimensionless function of the momentum transfer, it encodes how momentum is redistributed among the constituent partons, thereby providing a key probe of hadron structure. 

The first rigorous perturbative QCD (pQCD) predictions to the entire domain of large $Q^2$ processes date back to the 1980s \cite{Efremov:1979qk,Lepage:1980fj}. Since then, it has become widely accepted that the corresponding amplitudes are dominated by short-distance quark and gluon subprocesses. However, a controversy persisted regarding the relative roles of the Feynman mechanism and the hard-scattering mechanism, particularly in addressing the endpoint singularity in the momentum-fraction integration. This issue was not systematically resolved until the introduction of transverse momentum and the resummation of large logarithms \cite{Li:1992nu}, which together provided a consistent theoretical framework. Subsequently, efforts have been directed toward improving the accuracy of pQCD predictions by incorporating corrections from higher-order QCD loops as well as higher-twist light-cone distribution amplitudes (LCDAs) \cite{Li:2010nn,Cheng:2014gba,Hu:2012cp,Cheng:2015qra}. Recent advances include the next-to-next-to-leading-order (NNLO) calculation for the leading-twist LCDA contribution within the collinear factorization framework \cite{Chen:2023byr,Ji:2024iak}, and the next-to-leading-order (NLO) calculation up to subleading-twist LCDA contributions under the $k_T$ factorization scheme \cite{Cheng:2019ruz,Chai:2023htt}. 

Since the available experiments cannot access the genuine perturbative limit, the $k_T$ factorization framework, with its robust treatment of twist and power corrections, has a clear advantage over collinear factorization when dealing with exclusive processes in the intermediate-energy region. In this framework, the pion electromagnetic form factor is factorized as
\begin{equation}
{\cal F}_\pi(Q^2) = \sum_{t_i=2,3,4} \varphi^{t_1}(u_1,\mu_{r_1}) \, T_H^{t_i,{\rm LO+NLO}}(u_i,b_i,Q) S_t(u_i) \, e^{-S(u_i,b,Q)} \, \varphi^{t_2}(u_2,\mu_{r_2}). \label{eq:Fpi}
\end{equation}
Here $\varphi_{t_i}$ are the external pions' LCDAs \cite{Braun:2003rp,Braun:1989iv,Ball:2006wn}, with $t_i = 2,3,4$  being the twist subscript. The variables $u_i$ ($i=1,2$) are the longitudinal momentum fractions carried by the antiquark in the initial and final pions, respectively. The scales $\mu_{r_i}$ are the renormalization scales of the nonperturbative parameters. $T_H $ is the hard scattering kernel for the quark-gluon subprocess that has been studied at NLO. The large logarithms $\ln k_T^2$ are resummed into the Sudakov exponent $e^{-S}$ which highlights the hard transversal mechanism and selects the small $u_i \sim {\cal O}(0.1)$ regions in the momentum fraction integration. The residual logarithm $\ln x_i^2$, arising from the on-shell quark propagator, is resummed into the threshold factor $S_t$. 

In the conventioanl factorization formalism (\ref{eq:Fpi}), the threshold-suppressed hard scattering amplitude $T_H S_t$ encodes the hard interaction dynamics associated with both the longitudinal and transverse degrees of freedom. In contrast, Sudakov-multiplied LCDAs $e^{-S} \varphi_{t_i}$ capture only soft dynamics in the longitudinal direction, while transverse soft dynamics are absent. This omission arises because the renormalization scale $\mu_r$, whose inverse characterizes the small transverse separation between partons inside the pion, is usually chosen at the hard factorization scale in the pQCD approach. 

\begin{figure*}[h]
\centering
\sidecaption
\includegraphics[width=7cm,clip]{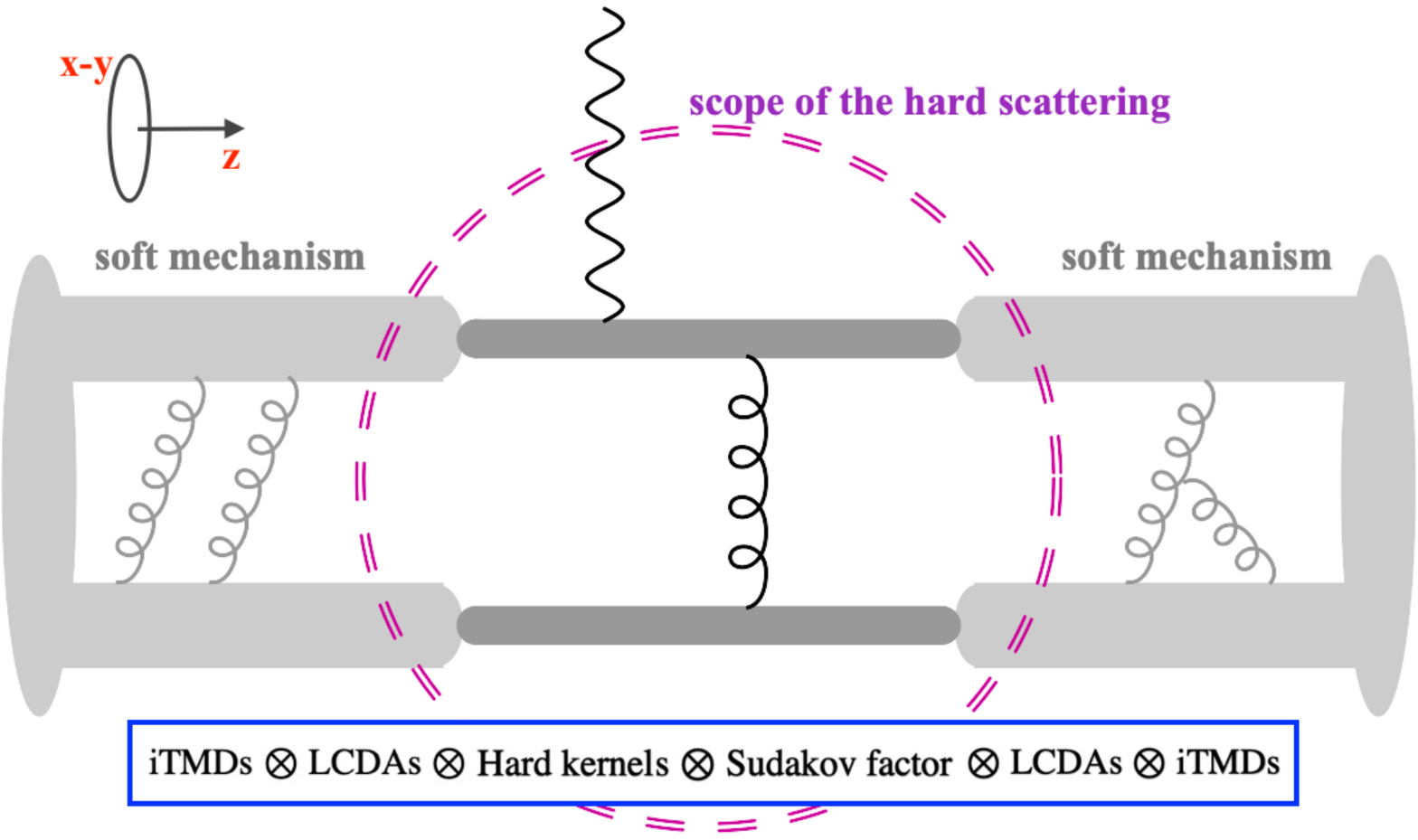}
\caption{The sketch map of pion electromagnetic form factors.}
\label{fig-1}       
\end{figure*}

Recently, we proposed introducing intrinsic transverse momentum distributions (iTMDs) to account for the missing soft dynamics and to complete the $k_T$ factorization formalism for exclusive processes \cite{Chai:2024tss,Chai:2025xuz}. Taking the pion electromagnetic form factor as an example, the sketch map of the complete factorization is shown in figure \ref{fig-1}. The strength of electromagnetic interactions is divided into two distinct regions by the double-dashed curves. Within the circle lies the realm of hard scattering (the central region of the electromagnetic potential field), described by the Sudakov-multiplied hard kernel ${\cal H} e^{-S}$. Outside the hard-scattering region, energetic pions move rapidly along the $z$ direction, accompanied by soft bremsstrahlung radiation. The bremsstrahlung radiation along the light cone is absorbed into the effects of higher-twist LCDAs. In the factorization formalism of Eq. (\ref{eq:Fpi}), the hard transverse dynamics of the LCDAs are defined by choosing the renormalization scale at the hard factorization scale, which aligns with the marginal region of the central potential field. In the exterior region, far from the central potential field, the soft radiation in the transverse plane (depicted by the thick light-gray stick) is notably absent from the definition of the LCDAs and has often been overlooked in previous pQCD studies. This constitutes the main motivation for our proposal. Then the Eq. (\ref{eq:Fpi}) is improved by replacing LCDAs by a soft pion wave fcuntion that is generally written by a product of LCDAs and iTMDs. 
\begin{equation}
\psi(u, k_T) = \frac{f_{\cal P}}{2\sqrt{6}} \varphi(u,\mu_r) \Sigma(u, k_T), 
\end{equation}
We adopt a simple Gaussian function, by considering a transversal harmonic oscillator with the transversal size parameter $\beta^2$, to parametrize them while preserving rotational invariance \cite{Jakob:1993iw,Kroll:2010bf}. 
\begin{equation}
\Sigma(u, k_T) = 16 \pi^2 \frac{\beta^2}{u (1-u)} e^{ - \frac{\beta^2 k_T^2}{u (1-u)} }. \label{eq:TMD-lt-kt} 
\end{equation}

\section{Form factors of light pseudoscalar mesons} \label{sec-1}

We would like to emphasize that the dominant contribution to the pion electromagnetic form factor arises from the twist-3 LCDAs owing to the chiral mass enhancement effect. Therefore, the first issue prior to the pQCD calculation is to determine the chiral mass with high accuracy. To this end, we employ the modular dispersion relation \cite{Cheng:2020vwr}, which relates the spacelike form factor to the timelike one without introducing any resonant model dependence. By taking the precise measurements of timelike form factors in the resonant regions \cite{BaBar:2012bdw,Belle:2008xpe,BESIII:2015equ} and the high-precision pQCD calculation in the large momentum transfer region as inputs, we extract the chiral mass $m_0^\pi = 1.84 \pm 0.07$ GeV. This value is consistent with the result of $1.79$ GeV obtained from chiral perturbation theory \cite{Leutwyler:1996qg}. The transversal size parameter $\beta_\pi^2 = 0.51 \pm 0.04$ GeV$^{-2}$ is constrained by the asymptotic behavior of the double-photon transition $\pi \to \gamma \gamma$. 

\begin{figure*}[h]
\centering        
\includegraphics[width=6cm,clip]{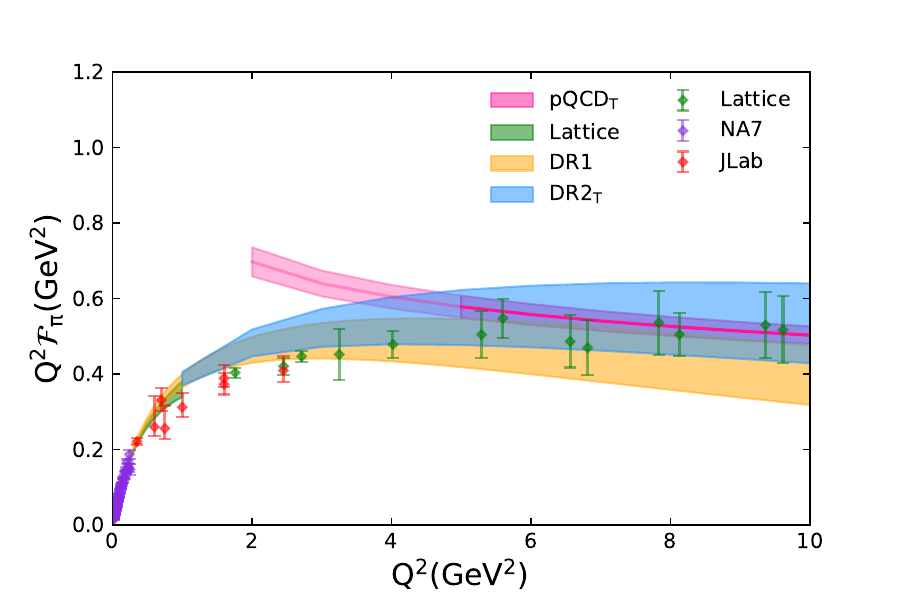} \hspace{2mm}
\includegraphics[width=6cm,clip]{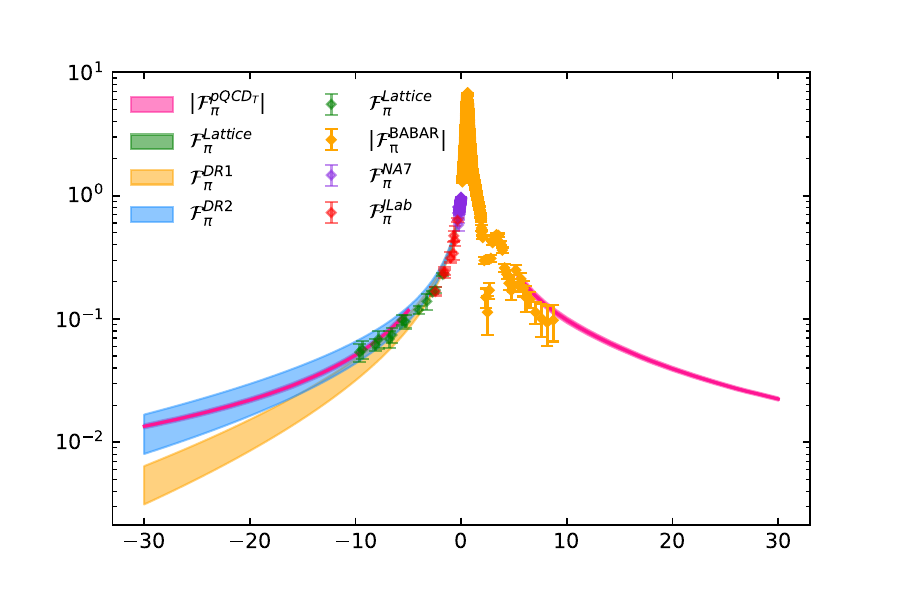} \vspace{-2mm}
\caption{Pion electromagnetic form factors in the spacelike region \cite{NA7:1986vav,JeffersonLabFpi-2:2006ysh,JeffersonLab:2008jve}(left) and the entire kinematic region (right). The magenta band shows the iTMDs-improved pQCD predictions; the yellow and blue bands in the spacelike region are the results from the modular dispersion relation without and with the high-energy tail on the timelike side, respectively.}
\label{fig-2} \end{figure*}

Figure \ref{fig-2} shows the iTMDs-improved pQCD calculation of the pion electromagnetic form factor. The result in the spacelike region agrees well with the lattice QCD evaluation \cite{Wang:2020nbf,Ding:2024lfj}, and the result in the timelike region exhibits an evolutionary trend compatible with electron-positron annihilation experiments in the intermediate energy region. We would like to highlight that the chiral enhancement effect from subleading-twist LCDAs is indispensable in generating the scaling behavior of the form factor, as seen in the lattice QCD evaluation, in the intermediate momentum ($\sim 5$ GeV$^2$) transfer region. Recently, the similar scaling behavior is also produced for the proton's Dirac form factor \cite{Yu:2025jcs}. 

\begin{figure*}[h] \centering
\includegraphics[width=6cm,clip]{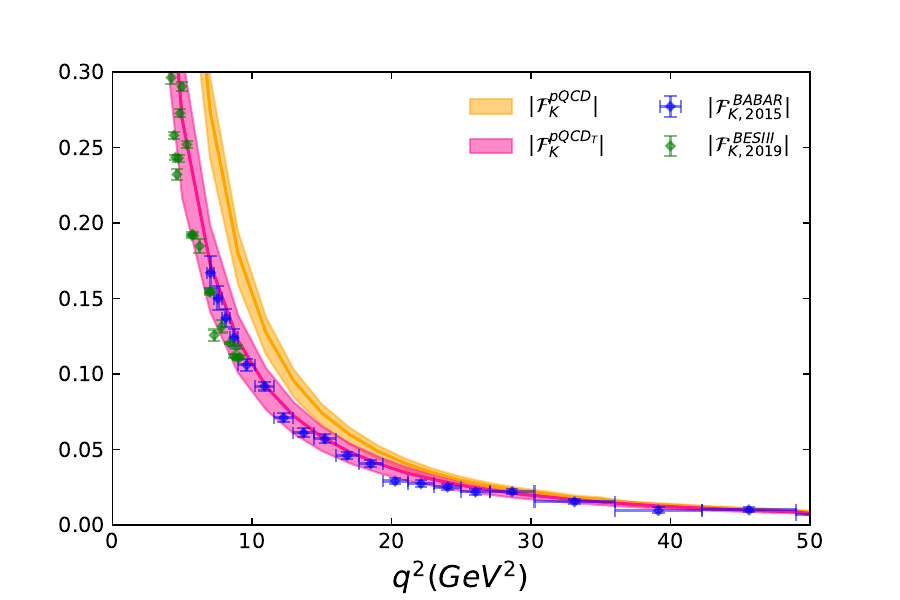} \hspace{2mm}
\includegraphics[width=6cm,clip]{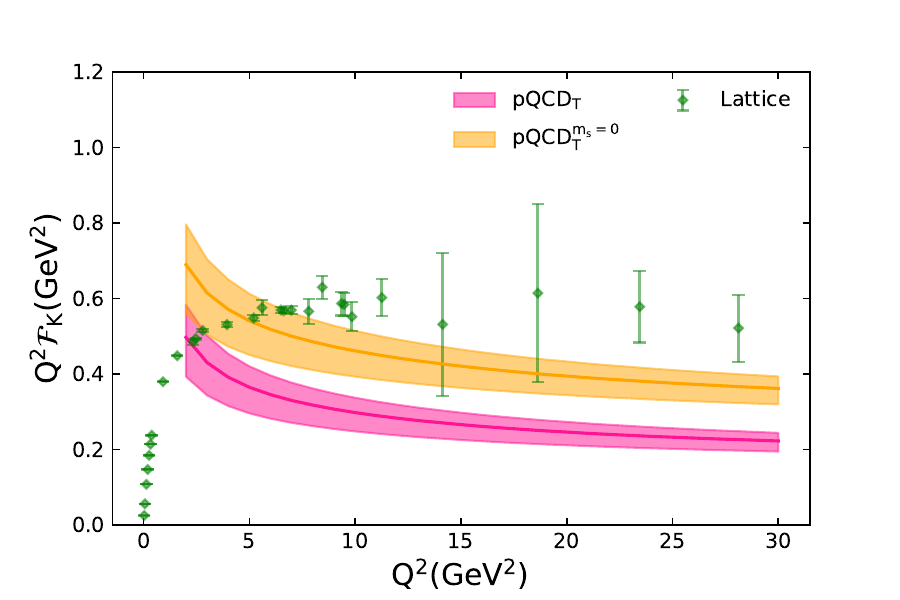} \vspace{-2mm}
\caption{Kaon electromagnetic form factors in the timelike (left) and spacelike regions (right).}
\label{fig-3} \end{figure*}

Figure \ref{fig-3} presents the iTMDs-improved pQCD prediction for the kaon electromagnetic form factor. In this analysis, we adopt the chiral mass of the kaon meson, $m_0^K = 1.90 \pm 0.09$ GeV derived from the well-known chiral perturbative theory relation without invoking light quark masses \cite{Leutwyler:1996qg}. The transverse size parameter, $\beta_K^2=0.30 \pm 0.05$ GeV$^2$, is extracted by fitting the pQCD calculation to the precise timelike measurements \cite{BaBar:2013jqz,BaBar:2015lgl,BESIII:2018ldc}, as shown in the left panel. The soft transverse dynamics incorporated in the iTMDs framework proves essential for explaining the precise timelike kaon form factor data. However, the corresponding predictions for the spacelike form factor, based on factorization approaches, are considerably smaller than the lattice results \cite{Ding:2024lfj}. We suggest that this discrepancy may arise from SU(3) flavor breaking, which introduces an additional ${\cal O}(m_s)$ term in the twist-3 LCDAs of the kaon meson. 

\begin{figure*}[h] \centering
\includegraphics[width=6cm,clip]{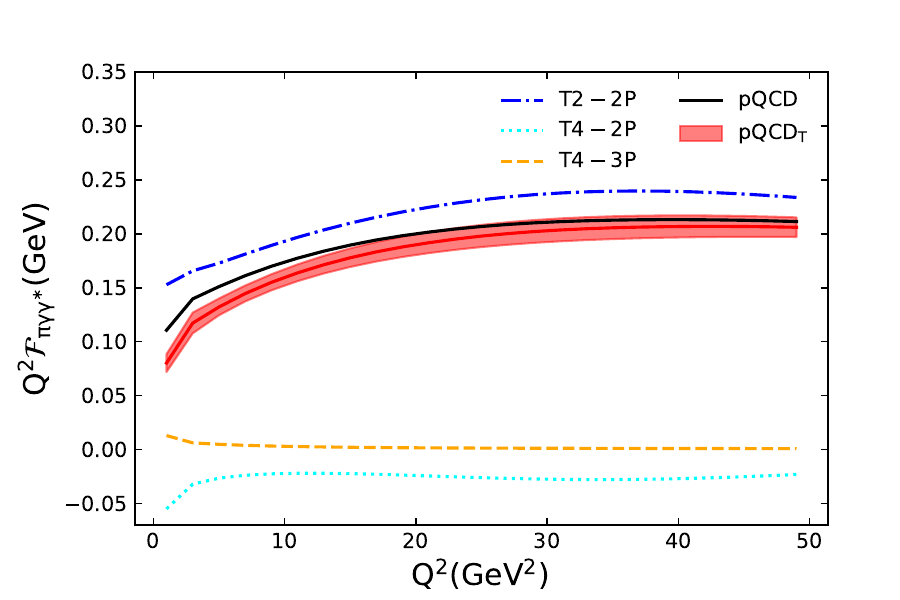} \hspace{2mm}
\includegraphics[width=6cm,clip]{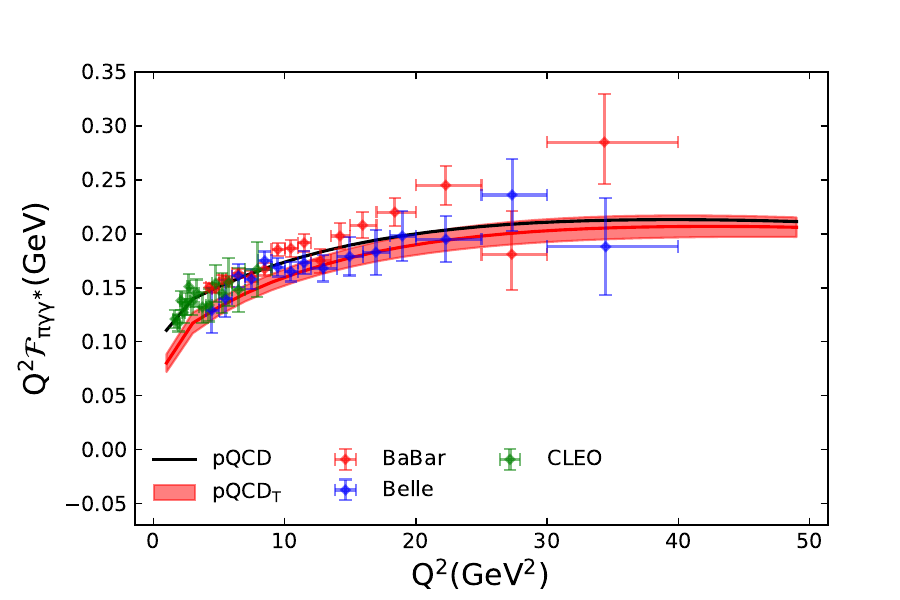} \vspace{-2mm}
\caption{Anatomy of the pion transition form factor in terms of twist contributions (left), compared with experimental measurements \cite{CLEO:1997fho,BaBar:2009rrj,Belle:2012wwz} (right).}
\label{fig-4} \end{figure*}

Using the extracted iTMDs associated with the LCDAs, we present the pion transition form factor in Figure \ref{fig-4}. The contributions from different twist LCDAs are explicitly shown in the left panel, where we find that the transition form factor is dominated by the leading-twist piece, since the twist-3 terms vanish due to the chiral symmetry preserved by the electromagnetic current conservation, and the subleading contribution starts at twist-4, which is highly suppressed relative to the leading one. The pQCD predictions are compared with the measurements from the BaBar and Belle collaborations in the right panel. Several features are evident. First, the iTMDs effect is significant in the small-to-intermediate $Q^2$ region (a few GeV$^2$), while it does not play a sizable role in the intermediate region. Second, the iTMDs-improved pQCD prediction supports the asymptotic QCD prediction $Q^2 F_{\pi\gamma\gamma^\ast} \to \sqrt{2} f_\pi$, which is more consistent with the Belle measurement. 

\begin{figure*}[h]\centering
\includegraphics[width=6cm,clip]{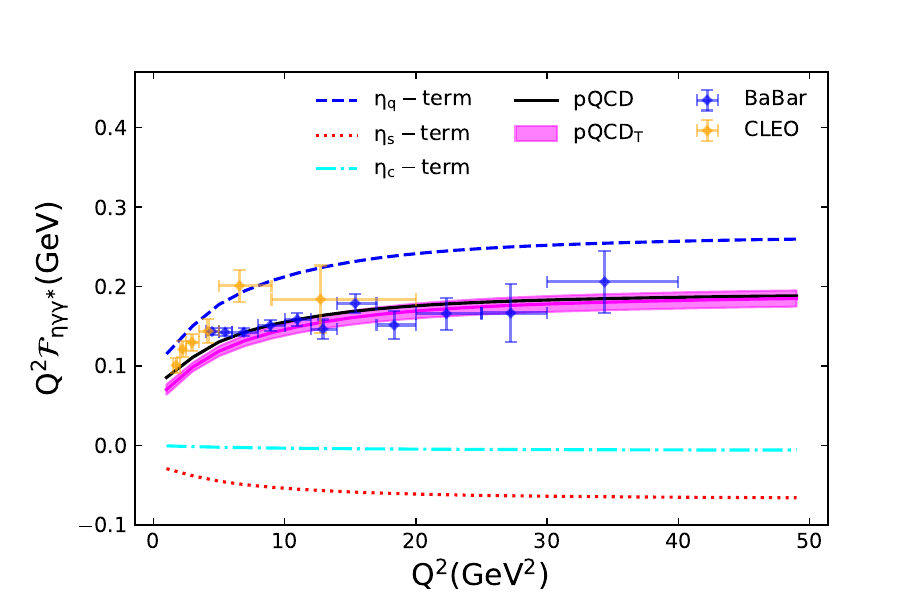} \hspace{2mm}
\includegraphics[width=6cm,clip]{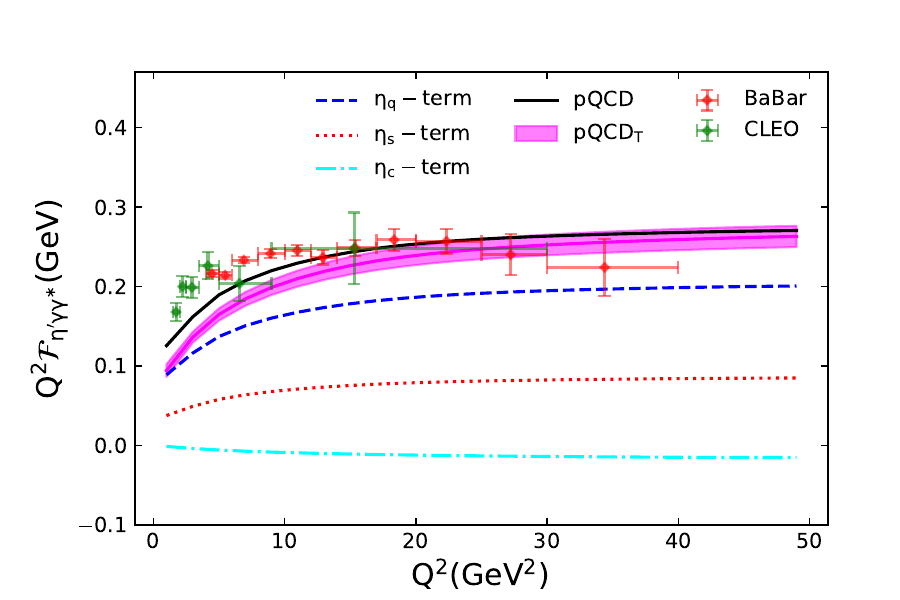} \vspace{-2mm}
\caption{Transition form factors of $\eta, \eta^\prime$ mesons}
\label{fig-5} \end{figure*}

The transition form factors of the isoscalar $\eta$ and $\eta^\prime$ mesons can be decomposed in terms of the orthogonal flavor basis $\eta_q, \eta_s$ and $\eta_c$. Neglecting the ${\cal O}(\theta_c^2)$ terms in the three-flavor basis mixing \cite{Feldmann:1998vh}, and taking $\beta_{\eta_q} = \beta_\pi, \beta_{\eta_s} = \beta_K$ and $\beta_{\eta_c} = 0.19$ GeV$^2$ from the two-photon decay width $\Gamma(\eta_c \to \gamma\gamma)$, we present the $\eta$ and $\eta^\prime$ transition form factors in Figure \ref{fig-5}, where the contributions from each flavor basis are also shown. We find that the current measurements favor a small mixing angle $\phi = \left(37.7 \pm 0.7\right)^\circ$ \cite{Cao:2012nj,Bali:2021qem} within our pQCD framework. 

\begin{figure}[h]
\centering\sidecaption
\includegraphics[width=6cm,clip]{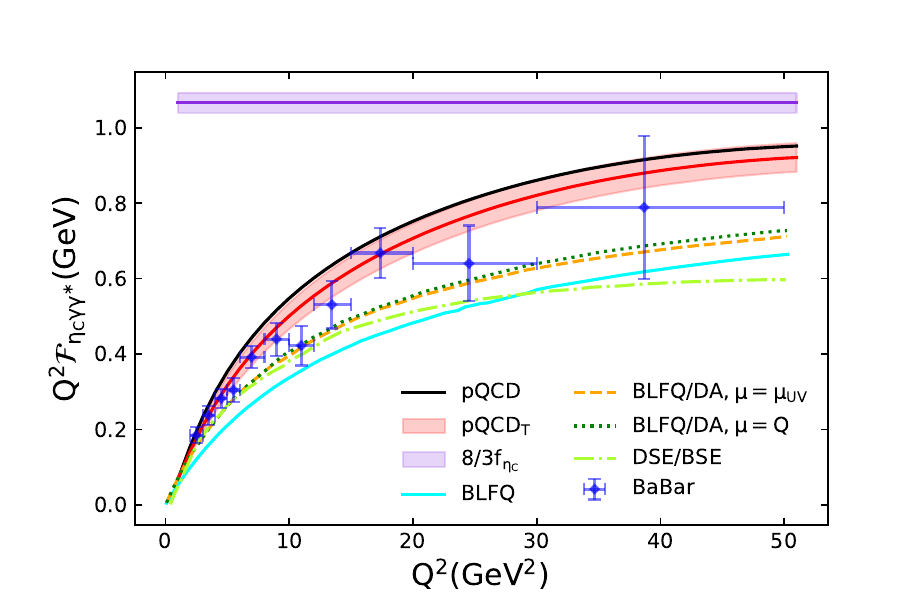} 
\caption{Transition form factors of $\eta_c$ state.}
\label{fig-6} \end{figure}

We separately present the $\eta_c$ form factor in figure \ref{fig-6}, where various theoretical predictions are shown together with the BaBar data \cite{BaBar:2010siw}. We note that our iTMD-improved pQCD prediction, calculated at leading order with leading-twist LCDAs, favors a larger value of 
${\cal F}_{\eta_c\gamma\gamma^\ast}(0) = 0.092$ GeV$^{-1}$. This explains why our curve lies slightly above those from light-front quantization (BLFQ) and the Bethe-Salpeter equation (BSE) \cite{Li:2021ejv,Chen:2016bpj}, where the form factors were normalized to $0.068$ GeV$^{-1}$ from the $\Gamma(\eta_c \to \gamma\gamma)$ data. This larger value has recently been confirmed by the BESIII measurement of $\Gamma(\eta_c \to \gamma\gamma)$ \cite{BESIII:2024rex}. 

\section{Summary}\label{sec-2}

We systematically study the electromagnetic and transition form factors of light pseudoscalar mesons ($\pi, K, \eta^{(\prime)}$) in perturbative QCD using $k_T$ factorization. Predictions include NLO QCD corrections (${\cal O}(\alpha_s^2)$) for leading and subleading twist LCDAs, and LO (${\cal O}(\alpha_s)$) for twist-four LCDAs. To capture soft transverse momentum effects, we introduce iTMDs, complementing the longitudinal LCDAs. This is motivated by hadron formation outside the hard interaction region where the sudakov suppression is weak. 

\begin{itemize}
    \item[$\bullet$] Higher-twist contributions, especially subleading-twist LCDAs, drive the observed scaling behavior of form factors at intermediate momentum transfers. 
    \item[$\bullet$] iTMDs are essential for describing the form factor lineshape at small-to-intermediate $Q^2$ , extending pQCD calculations down to a few GeV$^2$.
    \item[$\bullet$] iTMDs-improved predictions for pion and kaon electromagnetic form factors agree well with data. Discrepancies with lattice results of the spacelike kaon form factor suggest SU(3) breaking effects, offering a precision test of the ${\cal O}(m_s)$ term in twist-3 LCDAs.
    \item[$\bullet$] Including iTMDs, our predictions favor Belle data for the pion transition form factor and a small angle $\phi = \left( 37.7 \pm 0.7 \right)^\circ$ between the flavor basis $\eta_q$ and $\eta_s$ for $\eta-\eta^\prime$ mixing. The iTMDs effect also increases the $\eta_c$ transition form factor normalization, which is subsequently confirmed by BESIII.
    \item[$\bullet$] For the proton's Dirac form factor, unexpectedly large higher-power contributions from subleading-twist LCDAs arise from endpoint enhancement. We highlight that this enhancement, tamed by the $k_T$ resummation, is crucial for the approximate scaling of $Q^4 F_1(Q^2)$ data at intermediate $Q^2 \sim 10$ GeV$^2$. The study of iTMDs effect in proton form factors is in progress \cite{iTMDs-proton}.
\end{itemize}


\begin{thebibliography}{}

\bibitem{Mueller:1981sg}
A.~H.~Mueller, Perturbative QCD at High-Energies, Phys. Rept. \textbf{73}, 237 (1981). \url{doi:10.1016/0370-1573(81)90030-2}

\bibitem{PQCD}
S.J. Brodsky, G.P. Lepage, in: A.H. Mueller (Ed.), \textit{Perturbative Quantum Chromodynamics} (World Scientific, Singapore, 1989.)


\bibitem{Efremov:1979qk}
A.~V.~Efremov and A.~V.~Radyushkin, Factorization and Asymptotical Behavior of Pion Form-Factor in QCD, Phys. Lett. B \textbf{94}, 245-250 (1980). \url{doi:10.1016/0370-2693(80)90869-2}

\bibitem{Lepage:1980fj}
G.~P.~Lepage and S.~J.~Brodsky, Exclusive Processes in Perturbative Quantum Chromodynamics, Phys. Rev. D \textbf{22}, 2157 (1980). \url{doi:10.1103/PhysRevD.22.2157}

\bibitem{Li:1992nu}
H.~n.~Li and G.~F.~Sterman, The Perturbative pion form-factor with Sudakov suppression, Nucl. Phys. B \textbf{381}, 129-140 (1992). \url{doi:10.1016/0550-3213(92)90643-P}

\bibitem{Li:2010nn}
H.~n.~Li, Y.~L.~Shen, Y.~M.~Wang and H.~Zou, Next-to-leading-order correction to pion form factor in $k_T$ factorization, Phys. Rev. D \textbf{83}, 054029 (2011). \url{doi:10.1103/PhysRevD.83.054029}
\bibitem{Cheng:2014gba}
S.~Cheng, Y.~Y.~Fan and Z.~J.~Xiao, NLO twist-3 contribution to the pion electromagnetic form factors in $k_T$ factorization, Phys. Rev. D \textbf{89}, no.5, 054015 (2014). \url{doi:10.1103/PhysRevD.89.054015}
\bibitem{Hu:2012cp}
H.~C.~Hu and H.~n.~Li, Next-to-leading-order time-like pion form factors in $k_T$ factorization, Phys. Lett. B \textbf{718}, 1351-1357 (2013). \url{doi:10.1016/j.physletb.2012.12.006}
\bibitem{Cheng:2015qra}
S.~Cheng and Z.~J.~Xiao, Time-like pion electromagnetic form factors in $k_{T}$ factorization with the next-to-leading-order twist-3 contribution, Phys. Lett. B \textbf{749} (2015), 1-7. \url{doi:10.1016/j.physletb.2015.07.038}

\bibitem{Chen:2023byr}
L.~B.~Chen, W.~Chen, F.~Feng and Y.~Jia, Next-to-Next-to-Leading-Order QCD Corrections to Pion Electromagnetic Form Factors, Phys. Rev. Lett. \textbf{132}, no.20, 201901 (2024). \url{doi:10.1103/PhysRevLett.132.201901}
\bibitem{Ji:2024iak}
Y.~Ji, B.~X.~Shi, J.~Wang, Y.~F.~Wang, Y.~M.~Wang and H.~X.~Yu, Next-to-Next-to-Leading-Order QCD Prediction for the Pion Form Factor, Rev. Lett. \textbf{134}, no.22, 221901 (2025). \url{doi:10.1103/yx51-lc3q}

\bibitem{Cheng:2019ruz}
S.~Cheng, Pion and Kaon form factors in the perturbative QCD approach, Phys. Rev. D \textbf{100}, no.1, 013007 (2019). \url{doi:10.1103/PhysRevD.100.013007}
\bibitem{Chai:2023htt}
J.~Chai, S.~Cheng and J.~Hua, Study pion distribution amplitudes from the electromagnetic form factor by data-driven dispersion relation, Eur. Phys. J. C \textbf{83} (2023) no.7, 556. \url{doi:10.1140/epjc/s10052-023-11725-2}

\bibitem{Braun:2003rp}
V.~M.~Braun, G.~P.~Korchemsky and D.~M\"uller, The Uses of conformal symmetry in QCD, Prog. Part. Nucl. Phys. \textbf{51}, 311-398 (2003). \url{doi:10.1016/S0146-6410(03)90004-4}
\bibitem{Braun:1989iv}
V.~M.~Braun and I.~E.~Filyanov, Conformal Invariance and Pion Wave Functions of Nonleading Twist, Z. Phys. C \textbf{48}, 239-248 (1990). \url{doi:10.1007/BF01554472}
\bibitem{Ball:2006wn}
P.~Ball, V.~M.~Braun and A.~Lenz, Higher-twist distribution amplitudes of the K meson in QCD, JHEP \textbf{05}, 004 (2006). \url{doi:10.1088/1126-6708/2006/05/004}

\bibitem{Chai:2024tss}
J.~Chai and S.~Cheng, Shedding light on the intrinsic transversal momentum distributions of pions and kaons, Phys. Rev. D \textbf{111}, no.7, L071902 (2025). \url{doi:10.1103/PhysRevD.111.L071902}
\bibitem{Chai:2025xuz}
J.~Chai and S.~Cheng, Form factors of light pseudoscalar mesons from the perturbative QCD approach, JHEP \textbf{06}, 229 (2025). \url{doi:10.1007/JHEP06(2025)229}

\bibitem{Jakob:1993iw}
R.~Jakob and P.~Kroll, The Pion form-factor: Sudakov suppressions and intrinsic transverse momentum, Phys. Lett. B \textbf{315} (1993), 463-470 [erratum: Phys. Lett. B \textbf{319} (1993), 545]. \url{doi:10.1016/0370-2693(93)91642-Z}
\bibitem{Kroll:2010bf}
P.~Kroll, The form factors for the photon to pseudoscalar meson transitions - an update, Eur. Phys. J. C \textbf{71} (2011), 1623. \url{doi:10.1140/epjc/s10052-011-1623-4}

\bibitem{Cheng:2020vwr}
S.~Cheng, A.~Khodjamirian and A.~V.~Rusov, Pion light-cone distribution amplitude from the pion electromagnetic form factor, Phys. Rev. D \textbf{102}, no.7, 074022 (2020). \url{doi:10.1103/PhysRevD.102.074022}

\bibitem{BaBar:2012bdw}
J.~P.~Lees \textit{et al.} [BaBar], Precise Measurement of the $e^+ e^- \to \pi^+\pi^- (\gamma)$ Cross Section with the Initial-State Radiation Method at BABAR, Phys. Rev. D \textbf{86} (2012), 032013. \url{doi:10.1103/PhysRevD.86.032013}
\bibitem{Belle:2008xpe}
M.~Fujikawa \textit{et al.} [Belle], High-Statistics Study of the $\tau \to \pi^-\pi^0 \nu_\tau$ Decay, Phys. Rev. D \textbf{78} (2008), 072006. \url{doi:10.1103/PhysRevD.78.072006}
\bibitem{BESIII:2015equ}
M.~Ablikim \textit{et al.} [BESIII], Measurement of the $e^+e^- \to \pi^+\pi^-$ cross section between 600 and 900 MeV using initial state radiation, Phys. Lett. B \textbf{753} (2016), 629-638 [erratum: Phys. Lett. B \textbf{812} (2021), 135982]. \url{doi:10.1016/j.physletb.2015.11.043}

\bibitem{Leutwyler:1996qg}
H.~Leutwyler, The Ratios of the light quark masses, Phys. Lett. B \textbf{378}, 313-318 (1996). \url{doi:10.1016/0370-2693(96)00386-3}

\bibitem{NA7:1986vav}    
S.~R.~Amendolia \textit{et al.} [NA7], A Measurement of the Space - Like Pion Electromagnetic Form-Factor, Nucl. Phys. B \textbf{277} (1986), 168. \url{doi:10.1016/0550-3213(86)90437-2}
\bibitem{JeffersonLabFpi-2:2006ysh}
T.~Horn \textit{et al.} [Jefferson Lab F(pi)-2], Determination of the Charged Pion Form Factor at $Q^2 = 1.60$ and $2.45$ (GeV/c)$^2$, Phys. Rev. Lett. \textbf{97} (2006), 192001. \url{doi:10.1103/PhysRevLett.97.192001}
\bibitem{JeffersonLab:2008jve}
G.~M.~Huber \textit{et al.} [Jefferson Lab], Charged pion form-factor between $Q^2 = 0.60$ GeV$^2$ and 2.45 GeV$^2$. II. Determination of, and results for, the pion form-factor, Phys. Rev. C \textbf{78} (2008), 045203. \url{doi:10.1103/PhysRevC.78.045203}

\bibitem{Wang:2020nbf}
G.~Wang \textit{et al.} [$\chi$QCD], Lattice Calculation of Pion Form Factor with Overlap Fermions, Phys. Rev. D \textbf{104} (2021), 074502. \url{doi:10.1103/PhysRevD.104.074502}
\bibitem{Ding:2024lfj}
H.~T.~Ding, X.~Gao, A.~D.~Hanlon, S.~Mukherjee, P.~Petreczky, Q.~Shi, S.~Syritsyn, R.~Zhang and Y.~Zhao, QCD Predictions for Meson Electromagnetic Form Factors at High Momenta: Testing Factorization in Exclusive Processes, Phys. Rev. Lett. \textbf{133}, no.18, 181902 (2024). \url{doi:10.1103/PhysRevLett.133.181902}

\bibitem{Yu:2025jcs}
J.~X.~Yu, S.~Cheng, J.~J.~Han, H.~n.~Li and F.~S.~Yu, Tomography of high-twist proton structure through ep elastic scattering, Eur. Phys. J. C \textbf{86}, no.3, 300 (2026). \url{doi:10.1140/epjc/s10052-026-15554-x}

\bibitem{BaBar:2013jqz}
J.~P.~Lees \textit{et al.} [BaBar], Precision measurement of the $e^+e^- \to K^+K^-(\gamma)$ cross section with the initial-state radiation method at BABAR, Phys. Rev. D \textbf{88} (2013) no.3, 032013. \url{doi:10.1103/PhysRevD.88.032013}
\bibitem{BaBar:2015lgl}
J.~P.~Lees \textit{et al.} [BaBar], Study of the $e^+e^-\to K^+K^-$ reaction in the energy range from 2.6 to 8.0 GeV, Phys. Rev. D \textbf{92} (2015) no.7, 072008. \url{doi:10.1103/PhysRevD.92.072008}
\bibitem{BESIII:2018ldc}
M.~Ablikim \textit{et al.} [BESIII], Measurement of $e^{+} e^{-} \rightarrow K^{+} K^{-}$ cross section at $\sqrt{s} = 2.00 - 3.08$ GeV, Phys. Rev. D \textbf{99} (2019) no.3, 032001. \url{doi:10.1103/PhysRevD.99.032001}

\bibitem{CLEO:1997fho}
J.~Gronberg \textit{et al.} [CLEO], Measurements of the meson-photon transition form-factors of light pseudoscalar mesons at large momentum transfer, Phys. Rev. D \textbf{57} (1998), 33-54. \url{doi:10.1103/PhysRevD.57.33}
\bibitem{BaBar:2009rrj}      
B.~Aubert \textit{et al.} [BaBar], Measurement of the $\gamma \gamma^\ast \to \pi^0$ transition form factor, Phys. Rev. D \textbf{80} (2009), 052002. \url{doi:10.1103/PhysRevD.80.052002}
\bibitem{Belle:2012wwz}
S.~Uehara \textit{et al.} [Belle], Measurement of $\gamma \gamma^* \to \pi^0$ transition form factor at Belle, Phys. Rev. D \textbf{86} (2012), 092007. \url{doi:10.1103/PhysRevD.86.092007}

\bibitem{Feldmann:1998vh}
T.~Feldmann, P.~Kroll and B.~Stech, Mixing and decay constants of pseudoscalar mesons, Phys. Rev. D \textbf{58} (1998), 114006. \url{doi:10.1103/PhysRevD.58.114006}
\bibitem{Cao:2012nj}
F.~G.~Cao, Determination of the $\eta$-$\eta^\prime$ mixing angle, Phys. Rev. D \textbf{85} (2012), 057501. \url{doi:10.1103/PhysRevD.85.057501}
\bibitem{Bali:2021qem}
G.~S.~Bali \textit{et al.} [RQCD], Masses and decay constants of the $\ensuremath{\eta}$ and $\ensuremath{\eta}'$ mesons from lattice QCD, JHEP \textbf{08}, 137 (2021). \url{doi:10.1007/JHEP08(2021)137}

\bibitem{BaBar:2010siw}
J.~P.~Lees \textit{et al.} [BaBar], Measurement of the $\gamma \gamma^\ast \to \eta_c$ transition form factor, Phys. Rev. D \textbf{81} (2010), 052010. \url{doi:10.1103/PhysRevD.81.052010}
\bibitem{Li:2021ejv}
Y.~Li, M.~Li and J.~P.~Vary, Two-photon transitions of charmonia on the light front, Phys. Rev. D \textbf{105} (2022) no.7, L071901. \url{doi:10.1103/PhysRevD.105.L071901}
\bibitem{Chen:2016bpj}
J.~Chen, M.~Ding, L.~Chang and Y.~x.~Liu, Two Photon Transition Form Factor of $\bar{c}c $ Quarkonia, Phys. Rev. D \textbf{95} (2017) no.1, 016010. \url{doi:10.1103/PhysRevD.95.016010}
\bibitem{BESIII:2024rex}
M.~Ablikim \textit{et al.} [BESIII], Observation of the charmonium decay $\eta_c\to\gamma\gamma$ in $J/\psi\to\gamma\eta_c$, Phys. Rev. Lett. \textbf{134}, 181901 (2025). \url{doi:10.1103/PhysRevLett.134.181901}

\bibitem{iTMDs-proton}
J.~X.~Yu, J.~Chai, S.~Cheng and et.al., Unraveling Proton Structure through Electromagnetic Form Factors, in progress 

\end{thebibliography}
%
%

\end{document}